\documentclass[11pt]{iopart}
\usepackage{graphicx}

\begin{document}
\title[Imprints of the nuclear symmetry energy on the tidal deformability of neutron stars]{Imprints of the
nuclear symmetry energy on the tidal deformability of neutron stars}

\author{Plamen G. Krastev$^1$ and Bao-An Li$^2$}

\address{$^1$ Harvard University, Faculty of Arts and Sciences, Research Computing, 38 Oxford Street, Cambridge, MA 02138, U.S.A.}
\address{$^2$ Department of Physics and Astronomy, Texas A\&M University-Commerce, P.O. Box 3011, Commerce, TX 75429, U.S.A.}
\eads{\mailto{plamenkrastev@fas.harvard.edu}, \mailto{Bao-An.Li@tamuc.edu}}

\begin{abstract}
Applying an equation of state (EOS) with its symmetric nuclear matter (SNM) contribution and low-density symmetry energy $E_{sym}(\rho)$
constrained by heavy-ion reaction data, we calculate the dimensionless tidal deformability $\Lambda$ of neutron stars in coalescing
binary systems. Corresponding to the partially constrained EOS that previously predicted a radius of 11.5 km $\leq R_{1.4} \leq$ 13.6 km
for canonical neutron-star configurations, $\Lambda$ is found to be in the range  of 292 $\leq\Lambda_{1.4}\leq$ 680, consistent with
the very recent observation of the GW170817 event. We investigate the effect of the high-density behavior of $E_{sym}(\rho)$ on the tidal
properties of neutron stars and find that while $\Lambda$ depends strongly on the details of the symmetry energy, different trends of
$E_{sym}(\rho)$ lead to very similar values of $\Lambda$. In particular, the transition from stiff/soft to soft/stiff
$E_{sym}(\rho)$ could yield the same $\Lambda$. Thus, measuring $\Lambda$ alone may not determine completely the density dependence of
the symmetry energy. Coherent analyses of the dense neutron-rich nuclear matter EOS underlying both nuclear laboratory experiments
and astrophysical observations are therefore necessary to break this degeneracy and determine precisely the details of
the $E_{sym}(\rho)$.
\end{abstract}


\submitto{\jpg}


\section{Introduction}

The very first gravitational wave detection, GW170817, from a binary neutron-star merger~\cite{TheLIGOScientific:2017qsa,GBM:2017lvd,Abbott:2018exr},
together with its electromagnetic (EM)  counterpart, AT2017gfo (see, e.g., Ref.~\cite{Radice:2017lry} and references therein),
marks the beginning of the new era of {\it multi-messenger} astronomy, and already started to provide important insights for
astrophysics, cosmology, production of heavy elements, and nature of dense matter. Because gravity interacts extremely
weakly with matter, gravitational waves (GWs) could deliver detailed information about the neutron  star structure and
the underlying EOS of dense matter that is unaccessible with conventional astronomical observations. The connection
between the gravitational wave signal and the neutron star  EOS is provided by the tidal distortion of the  star in the
extremely strong gravitational field of its companion during the  inspiral~\cite{Hinderer:2009ca}. Although the  influence
of the EOS on the waveform is most significant during the later  stages of the inspiral and merger of two neutron  stars,
it has been pointed out~\cite{Flanagan:2007ix} that tidal deformation effects could be potentially measurable at earlier
times of the binary dynamics when the gravitational wave signal is relatively clean. These effects are quantified  in terms
of a single parameter -- the tidal deformability, $\lambda$, which characterizes the quadruple deformation of  the star in the
tidal gravitational field of its companion (see, e.g., Ref.~\cite{Shibata:2015}).

While earlier studies have shown that tidal effects modify the waveform only at the end of an inspiral, and their
signature may not be distinguishable from its point-particle post Newtonian
shape \cite{Hinderer:2009ca,1992ApJ...398..234K,Bildsten:1992,Lai:1996}, the joint detection of the GW170817 signal
by the LIGO and Virgo observatories has demonstrated clearly the measurability of tidal effects and placed
a constraint on the tidal deformability of neutron stars~\cite{TheLIGOScientific:2017qsa}.
With the projected sensitivities of later generation ground-based detectors, such as the Advanced LIGO~\cite{Aasi:2014mqd},
Virgo~\cite{TheVirgo:2014hva} and the Kamioka Gravitational Wave Detector  (KAGRA)~\cite{Aso:2013eba,Dominik:2014yma},
it is expected these effects to be observed in details  in the near future. In addition, with the rapid advancement  of the detection
technologies and the planned new generation gravitational wave observatories, it becomes therefore timely and important to provide
reliable estimates of the tidal deformation effects expected in inspiraling binary neutron stars. These estimates
play an important role in the overall framework for extracting information on the underlying neutron - star EOS, and
together with waveform templates constructed with the help of state-of-the-art numerical simulations (see, e.g.,
Refs.~\cite{Shibata:2015,Haas:2016cop}) and novel analytic techniques~\cite{Hinderer:2016eia}, could provide
valuable guidance for both the detection and interpretation of the gravitational wave signals from current and future detectors.

The response of a neutron star to an applied tidal gravitational field is quantified by the tidal deformability $\lambda$,
which depends on the details of the EOS of dense neutron-rich matter. In cold neutron star matter, the nucleonic part of the
EOS can be written in terms of the energy per nucleon as
\begin{eqnarray}\label{Eq.1}
E_n(\rho,\delta)\approx E_0(\rho)+E_{\rm{sym}}(\rho)\delta^2,
\end{eqnarray}
where $E_0(\rho)=E_n(\rho,0)$ is the energy per nucleon of symmetric nuclear matter (SNM), $E_{\rm{sym}}(\rho)$ is the
symmetry energy, and $\delta=(\rho_{\rm{n}}-\rho_{\rm{p}})/\rho$ is the isospin asymmetry, with $\rho_{\rm{n}}$, $\rho_{\rm{p}}$
and $\rho = \rho_{\rm{n}} + \rho_{\rm{p}}$, the neutron, proton and total density respectively. Expanding $E_0(\rho)$ up
to third order and $E_{sym}(\rho)$ up to second order around the saturation density $\rho_0$ we obtain:
\begin{equation}\label{Eq.1a}
E_0(\rho)\approx E_0(\rho_0) + \frac{K_0}{2}\left(\frac{\rho-\rho_0}{3\rho_0}\right)^2
+ \frac{J_0}{6}\left(\frac{\rho-\rho_0}{3\rho_0}\right)^3
\end{equation}
and
\begin{eqnarray}\label{Eq.1b}
E_{sym}(\rho) \approx E_{sym}(\rho_0) + L \left(\frac{\rho-\rho_0}{3\rho_0}\right)
+ \frac{K_{sym}}{2}\left(\frac{\rho-\rho_0}{3\rho_0}\right)^2.
\end{eqnarray}
The incompressibility $K_0$, the skewness coefficient $J_0$, the symmetry energy slope $L$, and the symmetry energy
curvature $K_{sym}$ evaluated at $\rho_0$ are defined in, e.g., Ref. \cite{Vidana:2009is}. Presently, the EOS of cold
nuclear matter under extreme conditions of density, pressure and/or isospin asymmetry still remains rather uncertain and
theoretically controversial, in particular at supra - saturation densities. Besides the tight constraint provided by the
maximum mass of neutron stars, extensive analyses of experimental data of heavy-ion reactions from intermediate to
relativistic energies, especially various forms of nucleon collective flow and the kaon production, have constrained
reasonably tightly the EOS of SNM up to about $4.5\rho_0$, see, e.g., Ref.~\cite{Danielewicz:2002pu}.
However, at higher densities the EOS remains largely uncertain mainly due to the poorly known high-density behavior of
the nuclear symmetry energy $E_{sym}(\rho)$, see, e.g., Refs. \cite{Li:1997px,ibook01,LP01,Lattimer:2004pg,Bar05,Ste05,EPJA}.
Besides astrophysical observations, both nuclear structure and reactions, especially with radioactive beams, provide useful
means to probe $E_{sym}(\rho)$ \cite{Bah14}. Indeed, thanks to the efforts and collaborations of both the nuclear physics
and astrophysics communities, significant progress has been made in recent years in constraining  the symmetry energy around
and below nuclear matter saturation density using results from both  astrophysical observations and terrestrial nuclear experiments,
see, e.g., Refs. \cite{LCK08,Tsa12,LiHan13,Hor14,Steiner14,Bal16}. Previously, several authors have studied neutron-star
tidal effects in inspiraling binaries and their detectability using both polytropic \cite{Hinderer:2007mb,Binnington:2009bb,Damour:2009vw}
and hadronic \cite{Hinderer:2009ca,Postnikov:2010yn,Moustakidis:2016sab,Kumar:2016dks} equations of state, and
investigated the influence of the EOS on the waveform. The GW170817 event has renewed the interest in the EOS of
dense neutron-rich matter and propelled an enormous amount of new detailed studies of various aspects of the EOS and its
applications to the physics of neutron stars and gravitational waves. Specifically, new constraints on the neutron-star radius and
the EOS of dense matter have been reported based on the extracted tidal deformability from the GW170817 observation. For instance,
the LIGO and Virgo collaborations have recently refined their analysis~\cite{Abbott:2018exr} and have reported the radii of the
neutron-star binary components in the range of 10.5 km $\leq R_{\{1,2\}} \leq$ 13.3 km. Other studies, based on various
many-body methods and nuclear interactions, have inferred a rather consistent upper limit on the canonical neutron-star radius
of $R_{1.4} \leq$ 13.7 km \cite{Zhou:2017pha,Annala:2017llu,Fattoyev:2017jql,Most:2018hfd,Raithel:2018ncd,Tews:2018chv,Malik:2018zcf,Lim:2018bkq}
using the original findings of Ref. \cite{TheLIGOScientific:2017qsa}.

In this work, applying an EOS with its SNM part and low-density $E_{sym}(\rho)$ constrained by heavy-ion reaction data, we
calculate the tidal deformability of neutron stars in coalescing binary systems and examine the effects of $E_{sym}(\rho)$
on $\lambda$. In particular, we investigate the impact of transition from stiff/soft to soft/stiff symmetry energy on the
tidal properties of neutron stars and discuss the implications of our results for the EOS of dense matter and the interpretation
of current  and future gravitation-wave signals from inspiraling neutron-star binaries.

This paper is organized as follows. After the introductory remarks in this section, in Sec. 2 we discuss the
necessary formalism to calculate the tidal deformability $\lambda$. In Sec. 3 we briefly review the main features
of the partially constrained EOS. We present our results for the tidal deformability of neutron stars and discuss
the effects of the EOS in Sec. 4. At the end, we conclude in Sec. 5 with a short summary and outlook of
future investigations.

{\it Conventions}: We use units in which $G = c = 1$.

\section{Formalism for calculating the neutron star tidal deformability}

In this section we briefly recall the formalism for calculating the neutron star tidal deformability $\lambda$.
As two neutron stars approach each other during the early stages of an inspiral they experience tidal deformation
effects quantified in terms of $\lambda$. This parameter is defined as~\cite{Hinderer:2009ca,Flanagan:2007ix,Damour:2009vw}
\begin{equation}\label{Eq.2}
\lambda = -\frac{Q_{ij}}{\mathcal{E}_{ij}},
\end{equation}
where $Q_{ij}$ is the induced mass quadruple moment of a star in the gravitational tidal field $\mathcal{E}_{ij}$
of its companion. The tidal deformability can be expressed in terms of the neutron star radius, $R$, and
dimensionless tidal Love number, $k_2$ as
\begin{equation}\label{Eq.3}
\lambda = \frac{2}{3}k_2 R^5.
\end{equation}
The tidal Love number $k_2$ is calculated using the following
expression~\cite{Hinderer:2007mb,Postnikov:2010yn,Moustakidis:2016sab}
\begin{eqnarray}\label{Eq.4}
k_2(\beta, y_R) && = \frac{8}{5}\beta^5(1-2\beta)^2 [2-y_R+2\beta(y_R-1)]                \nonumber \\
                          &&\times \{2\beta[6-3y_R+3\beta(5y_R-8)]                       \nonumber \\
                          &&+ 4\beta^3[13-11y_R + \beta(3y_R-2)                          \nonumber \\
                          &&+ 2\beta^2(1+y_R)] +3(1-2\beta)^2 [2-y_R                     \nonumber \\
                          &&+ 2\beta(y_R-1)]\ln(1-2\beta)\}^{-1},
\end{eqnarray}
where $\beta \equiv M / R$ is the dimensionless compactness parameter and $y_R \equiv y(R)$
is solution of the following first order differential equation
\begin{equation}\label{Eq.5}
\frac{dy(r)}{dr}=-\frac{y(r)^2}{r}-\frac{y(r)}{r}F(r)-rQ(r),
\end{equation}
with
\begin{eqnarray}
F(r)&& =\left\{1-4\pi r^2[\varepsilon(r)-p(r)]\right\}\left[1-\frac{2m(r)}{r}\right]^{-1},        \label{Eq.6} \\
Q(r)&& = 4\pi\left[5\varepsilon(r)+9p(r)+\frac{\varepsilon(r)+p(r)}{c_s^2(r)}
        -\frac{6}{r^2}\right]\left[1-\frac{2m(r)}{r}\right]^{-1}                                  \nonumber    \\
    && - \frac{4m^2(r)}{r^4}\left[1+\frac{4\pi{r^3}p(r)}{m(r)}\right]^2\left[1-\frac{2m(r)}{r}\right]^{-2}, \label{Eq.7}
\end{eqnarray}
where $c_s^2(r)\equiv dp(r)/d\varepsilon(r)$ is the squared speed of sound. Starting at the center of the star,
for a given EOS Eq.~(\ref{Eq.5}) needs to be integrated self-consistently together with the Tolman-Oppenheimer-Volkoff
equations, i.e.,
\begin{eqnarray}
\frac{dp(r)}{dr}&& = -\frac{\varepsilon(r)m(r)}{r^2}
\left[1+\frac{p(r)}{\varepsilon(r)}\right]\left[1+\frac{4\pi{r^3}p(r)}{m(r)}\right]
\left[1-\frac{2m(r)}{r}\right]^{-1},           \label{Eq.8} \\
\frac{dm(r)}{dr}&&=4\pi\varepsilon(r)r^{2}.    \label{Eq.9}
\end{eqnarray}
Imposing the boundary conditions at $r=0$ such that, $y(0)=2$, $m(0)=0$, and $p(0)=p_c$,
the Love number $k_2$ and the tidal deformability $\lambda$ can be readily calculated.
One can also compute the dimensionless tidal deformability $\Lambda$, which is related to
the compactness parameter $\beta$ and the Love number $k_2$ through
\begin{equation}
\Lambda = \frac{2}{3}\frac{k_2}{\beta^5}. \label{Eq.10}
\end{equation}

The total tidal effect of two neutron stars in an inspiraling binary system is given by the
mass-weighted (dimensionless) tidal deformability (see, e.g., Refs.~\cite{Hinderer:2009ca,Damour:2009vw})
\begin{equation}
\tilde{\Lambda} = \frac{16}{13}\frac{(M_1+12M_2)M_1^4\Lambda_1+(M_2+12M_1)M_2^4\Lambda_2}{(M_1+M_2)^5}, \label{Eq.11}
\end{equation}
where $\Lambda_1=\Lambda_1(M_1)$ and $\Lambda_2=\Lambda_2(M_2)$ are the (dimensionless) tidal
deformabilities of the individual binary components. As pointed out previously~\cite{Hinderer:2009ca},
although $\Lambda$ is calculated for single neutron stars, the universality of the neutron-star EOS
allows us to predict the tidal phase contribution for a given binary system from each EOS. For equal-mass
binary systems $\tilde{\Lambda}$ reduces to $\Lambda$.  The weighted (dimensionless) deformability
$\tilde{\Lambda}$ is usually plotted as a function of the chirp mass  $\mathcal{M}=(M_1 M_2)^{3/5}/M_T^{1/5}$
for various values of the asymmetric mass ratio  $\eta=M_1M_2/M_T^2$, where $M_T=M_1+M_2$ is the total mass of
the binary.

\section{Partially constrained equation of state of neutron-rich matter with the MDI (momentum-dependent interaction)}

The tidal deformability depends on the neutron-star EOS through both the tidal Love number $k_2$
and stellar radius $R$ (see Eq.~(\ref{Eq.3})). As already mentioned in the introduction, current
theoretical predictions of the nucleonic EOS diverge widely mainly because of the uncertain density
dependence of the nuclear symmetry energy, especially at high densities. In the ongoing efforts to
constrain the EOS and provide useful guidance to theoretical models, it is very important and
timely to determine what information could be extracted from precise measurements of $\Lambda$, and also
what aspects of the EOS could be better determined with the help of this information. This is clearly
a twofold problem. On the one hand, gravitational wave observations from inspiraling compact binaries
are expected to place more stringent constraints on the EOS, in particular on the high-density behavior
of the symmetry energy (see, e.g., Ref. \cite{Fattoyev:2012uu}). On the other hand, to extract useful
information on the details of the EOS from  gravitational waves, one needs reliable estimates of the
tidal deformation effects and accurate waveform  templates, which require reliable estimates of $\Lambda$.
To provide accurate estimates of the tidal deformability one needs to reduce the uncertainty of the
high-density $E_{sym}(\rho)$. Because the MDI EOS \cite{Das:2002fr,Li04} has its SNM part and symmetry
energy $E_{sym}(\rho)$ constrained by heavy-ion reaction data up to about $4.5\rho_0$ and $1.2\rho_0$
respectively, it would be interesting to compare the $\Lambda$ values using the MDI EOS with the recent
observational constraints from LIGO/Virgo \cite{TheLIGOScientific:2017qsa,GBM:2017lvd,Abbott:2018exr}.
In this study we assume a simple model of stellar matter of nucleons and light leptons
(electrons and muons) in beta-equilibrium.

\begin{figure}[t!]
\centering
\includegraphics[scale=0.5]{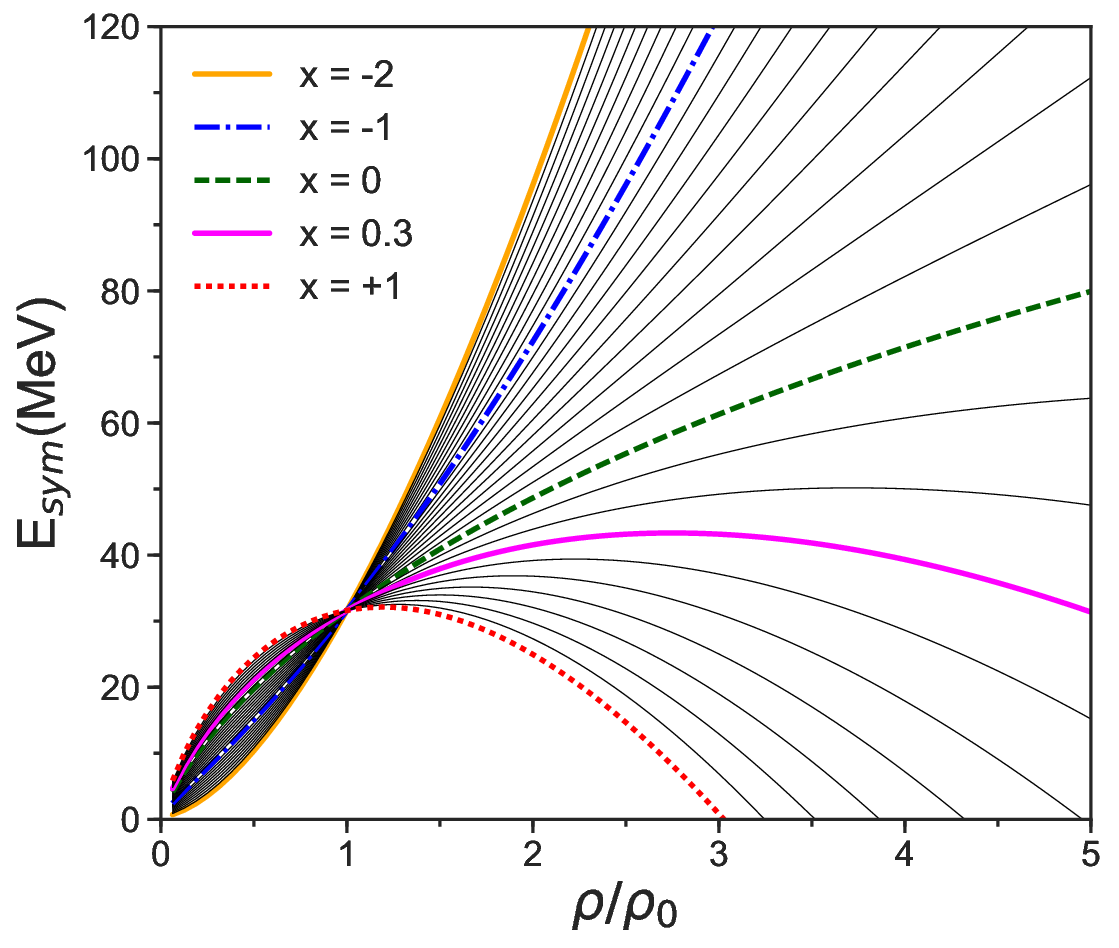}
\caption{(Color online) The density dependence of nuclear symmetry energy using the MDI interaction by varying the parameter $x$
which measures the relative strength of the effective three-body force in the isotriplet channel with respect to the isosinglet
channel. In the figure $x$ is varied from -2 to +1 in steps of $\Delta x = 0.1$.} \label{fig1}
\end{figure}

Below we briefly summarize the most relevant features of the MDI EOS. (For a comprehensive discussion  see,
e.g., Refs.~\cite{Das:2002fr,Li04}.) With the MDI interaction, the potential energy density $V(\rho,T, \delta )$
of an asymmetric nuclear matter at finite temperature $T$ is parameterized as~\cite{Das:2002fr,Li:2005jy}
\begin{eqnarray}
V(\rho,T,\delta ) &=&\frac{A_{u}(x)\rho _{n}\rho _{p}}{\rho _{0}}+\frac{A_{l}(x)}{
2\rho _{0}}(\rho _{n}^{2}+\rho _{p}^{2})+\frac{B}{\sigma +1}\frac{\rho^{\sigma +1}}{\rho _{0}^{\sigma }}  \nonumber \\
&\times &(1-x\delta ^{2})+\frac{1}{\rho _{0}}\sum_{\tau ,\tau ^{\prime
}}C_{\tau ,\tau ^{\prime }}  \nonumber \\
&\times &\int \int d^{3}pd^{3}p^{\prime }\frac{f_{\tau }(\vec{r},\vec{p},T
)f_{\tau ^{\prime }}(\vec{r},\vec{p}^{\prime
},T)}{1+(\vec{p}-\vec{p}^{\prime })^{2}/\Lambda ^{2}}, \label{Eq.12}
\end{eqnarray}
where $f_{\tau }(\vec{r},\vec{p},T)$ is the nucleon phase-space distribution function at coordinate
$\vec{r}$, momentum $\vec{p}$ and temperature $T$. The corresponding single-particle potential deduced
from the Hartree-Fock approach is given by
\begin{eqnarray}
&&U_{\tau }(\rho ,T,\delta ,\vec{p})=A_{u}(x)\frac{\rho _{-\tau
}}{\rho
_{0}}+A_{l}(x)\frac{\rho _{\tau }}{\rho _{0}}  \nonumber  \label{mdi} \\
&&+B\left( \frac{\rho }{\rho _{0}}\right) ^{\sigma }(1-x\delta ^{2})-8\tau x
\frac{B}{\sigma +1}\frac{\rho ^{\sigma -1}}{\rho _{0}^{\sigma
}}\delta \rho
_{-\tau }  \nonumber \\
&&+\sum_{t=\tau ,-\tau }\frac{2C_{\tau ,t}}{\rho _{0}}\int d^{3}\vec{p}
^{\prime }\frac{f_{t}(\vec{r},\vec{p}^{\prime
},T)}{1+(\vec{p}-\vec{p}^{\prime })^{2}/\Lambda ^{2}},  \label{Eq.13}
\end{eqnarray}
where $\tau =1/2$ ($-1/2$) for neutrons (protons). The parameters $x$, $A_{u}(x)$, $A_{\ell }(x)$, $B$,
$C_{\tau ,\tau }$,$C_{\tau ,-\tau }$, $\sigma$, and $\Lambda $ are fixed by saturation properties of SNM,
both isoscalar and isovector nucleon optical potentials, as well as a specified magnitude and slope of the
symmetry energy at saturation density of nuclear matter as discussed in detail in Ref.~\cite{Das:2002fr}.

\begin{figure}[t!]
\centering
\includegraphics[scale=0.6]{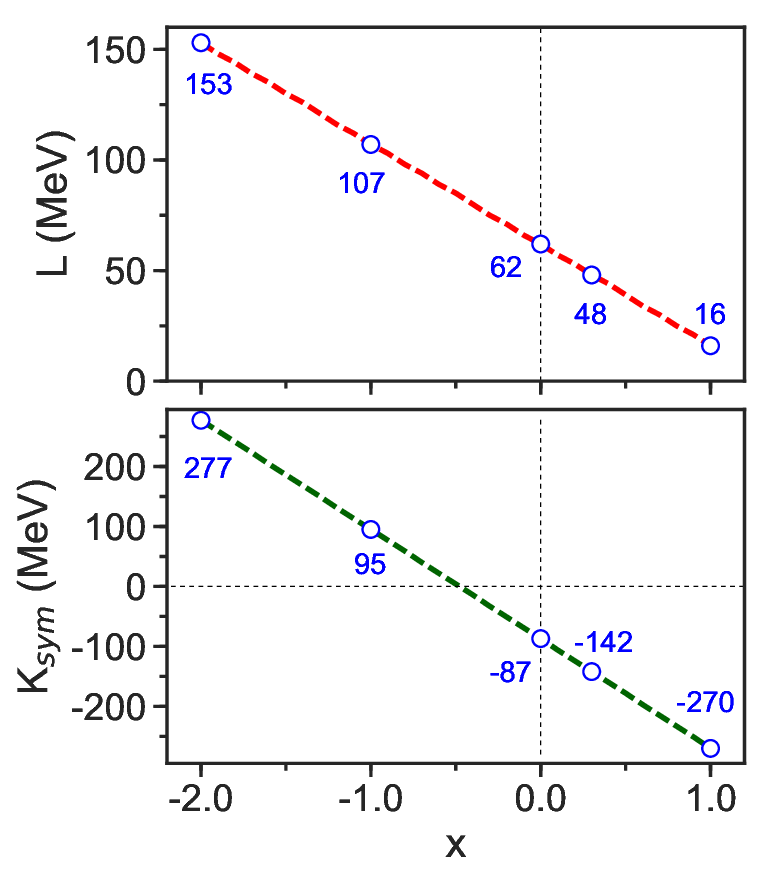}
\caption{(Color online) Slope of symmetry energy $L$ (upper panel) and curvature of symmetry energy $K_{sym}$ (lower panel)
evaluated at saturation density $\rho_0$ as functions of the parameter $x$. Numerical values are shown at
$x$ equal to -2, -1, 0, 0.3 and +1.} \label{fig2}
\end{figure}

The parameter $x$ in Eq.~(\ref{Eq.12}) was introduced to account for the uncertain spin-isospin dependence
of the three-body nuclear force in dense neutron-rich matter. It measures the relative strength of the isotriplet
to isosinglet interaction of the effective three-body force in the modified Gogny interaction. By varying the parameter
$x$, one can mimic the largely uncertain density dependence of the nuclear symmetry energy $E_{sym}(\rho)$ as
predicted by different nuclear many-body theories using various interactions. By design, different values of
$x$ can lead to widely different trends for the $E_{sym}(\rho)$ without changing the SNM EOS and the magnitude
of the symmetry energy at saturation density. This is illustrated in Fig.~\ref{fig1} where we show representative
examples of the $E_{sym}(\rho)$ for values of $x$ in the interval between -2 and +1. Here we need to emphasize that
various values of $x$ correspond to various values of $L$ and $K_{sym}$, i.e., varying $x$ changes
both parameters simultaneously as shown in Fig.~\ref{fig2}.

The MDI interaction has been used extensively in studies of heavy-ion reactions and neutron stars. It has also
been used to investigate thermodynamical properties of hot, dense and neutron-rich matter, see, e.g.,
Refs.~\cite{Xu1,Prakash}. Previously it has been demonstrated that only EOSs with values of $x$ in the range
between -1 and 0 have symmetry energy consistent with the terrestrial nuclear laboratory data \cite{Li:2005jy,Tsang04,Li:2005sr}.
Therefore these two limiting cases are expected to determine the boundaries of the most probable neutron star
configurations, and in turn the range of the most probable values of the (dimensionless) tidal deformability $\Lambda$.
For the purpose of this study, we consider EOSs with several representative values of $x$: -2, -1, 0, and 0.3.

\begin{figure}[t!]
\centering
\includegraphics[scale=0.5]{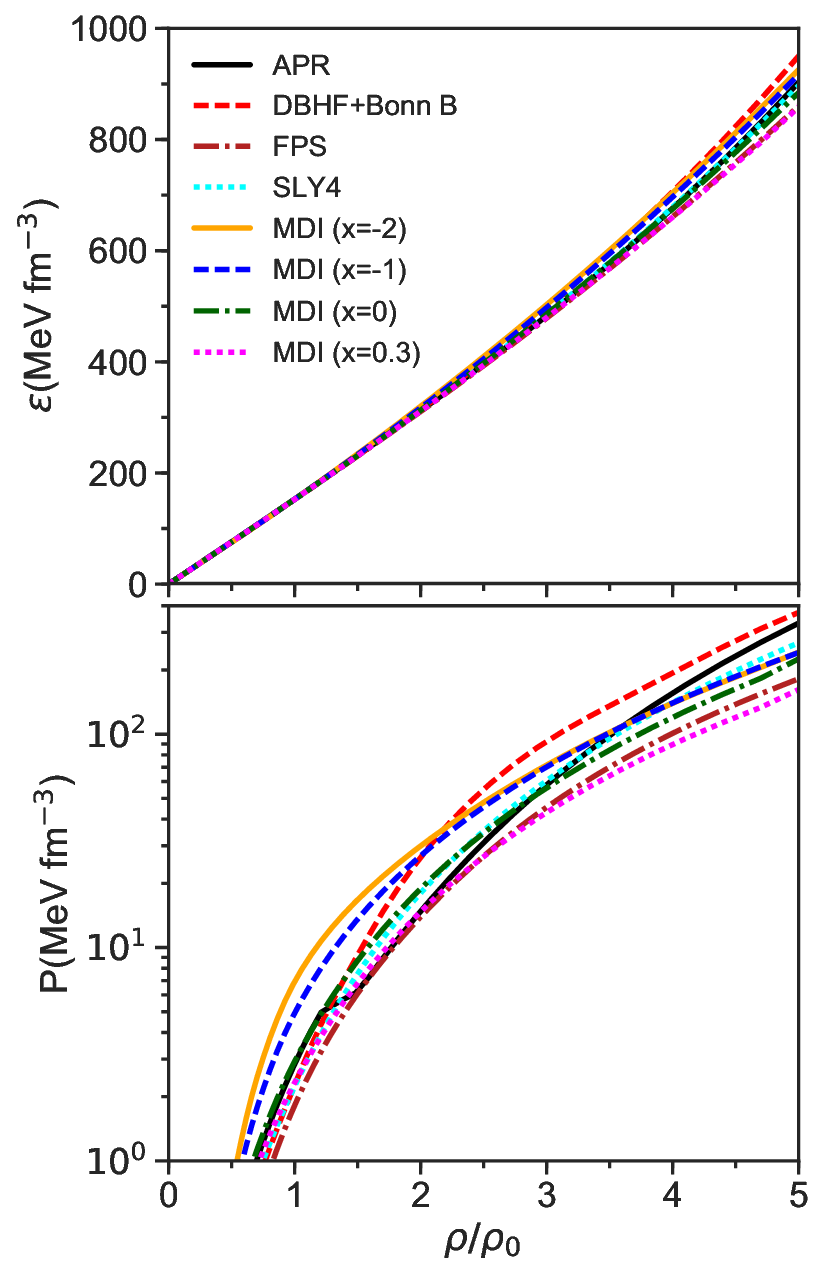}
\caption{(Color online) Equation of state. The upper frame shows the mass energy density as a function of
baryon density (in units of  saturation density, $\rho_0\approx 0.16fm^{-3} $) and the lower frame shows
the total pressure (including the lepton contributions) versus baryon density. (The "dip" exhibited by
the density curve of the APR EOS is due to a phase transition from low density phase (LDP) to high density
phase (HDP). See Akmal et al.~\cite{Akmal:1998cf} for details.)} \label{fig3}
\end{figure}

We show the EOSs applied in this work in Fig.~\ref{fig3}. The upper panel displays the total energy density,
$\varepsilon$, as a function of the baryon number density, and the lower panel shows the total pressure $P$.
In addition to the MDI EOS, we also include several EOSs frequently used in studies of neutron-star properties
and related phenomena. Namely, we also display results by Akmal et al.~\cite{Akmal:1998cf} with the $A18+\delta\upsilon+UIX*$
interaction (APR), Dirac-Brueckner-Hartree-Fock (DBHF) calculations~\cite{Alonso:2003aq,Krastev:2006ii} with Bonn B
One-Boson-Exchange (OBE) potential~\cite{Machleidt:1989} (DBHF+Bonn B), Pandharipande and Ravenhall~\cite{P&R:1989}
(FPS), and Douchin and Haensel~\cite{Douchin:2001sv} (SLY4). Below approximately $0.07fm^{-3}$ the EOSs
shown in Fig.~\ref{fig3} are supplemented by a crustal EOS, which is more suitable at lower densities. For the
purpose of this work we assume the same core-to-crust transition density for all MDI EOSs with various values
of $x$. For the inner crust we apply the EOS by Pethick et al.~\cite{PRL1995} and for the outer crust the one
by Haensel and Pichon \cite{HP1994}. At higher densities we assume a continuous functional for the equations
of state.

We emphasize that the EOS of SNM with the MDI interaction is constrained by the available data on collective
flow and kaon production in relativistic heavy-ion collisions~\cite{Danielewicz:2002pu} (see, e.g., Fig. 1 in
Ref.~\cite{Krastev:2008PLB}), while the $E_{sym}(\rho)$ is constrained around the saturation density by isospin
diffusion data ~\cite{Tsang04} to be between that with $x=-1$ and $x=0$ \cite{Li:2005jy,Li:2005sr}. Such a MDI EOS
constrained by the terrestrial laboratory data has been used in various calculations of neutron star properties
and astrophysical phenomena since about 2006. For example, it has been used to constrain the neutron star
radius~\cite{Li:2005sr} with an estimated range consistent with the observational data. More quantitatively,
corresponding to the parameter $x$ between 0 and -1, the MDI EOS predicted a radius of 11.5 km $< R_{1.4} < 13.6$ km
for a canonical neutron star of mass 1.4$M_{\odot}$~\cite{Li:2005sr}. This result was later found consistent with
estimates using the spin rate of the fastest pulsar PSR-J1748-2446ad \cite{Hes06}, thermonuclear bursts on neutron
star surfaces and spectra of neutron stars in quiescence~\cite{Steiner14}. Interestingly, the limits of
10.5 km $\leq R_{\{1,2\}} \leq$ 13.3 km at the 90\% credible level, for the component masses between 1.16 and
1.60 $M_{\odot}$, determined very recently by the refined analysis of the the GW170817 observation \cite{Abbott:2018exr}
are consistent with but less restrictive than the existing constraints on the radii of canonical neutron stars mentioned above.
Since the correlation between the radii of neutron stars and sizes of neutron-skins of heavy nuclei has been an interesting
topic in exploring the EOS of neutron-rich matter using the multi-messenger approach~\cite{chuck}, it is worth noting
that the size of neutron-skin in $^{208}$Pb was  predicted to be 0.22 fm and 0.28 fm with the MDI EOS of $x=0$ and
$x=-1$, respectively~\cite{Li:2005sr}. The constrained MDI EOS was also applied to study the possible time variation
of the gravitational constant $G$~\cite{Krastev:2007en} with the help of the {\it gravitochemical heating} approach
developed by Jofre et al.~\cite{Jofre:2006ug}. In addition, it was also used to limit a number of other global, transport
and thermal properties of both static and rapidly rotating neutron stars \cite{Krastev:2008PLB,KLW2,WKL:2008ApJ,Newton,Junxu}.

\section{Results and discussion}

\subsection{Love number and tidal deformability}

\begin{figure}[t!]
\centering
\includegraphics[scale=0.5]{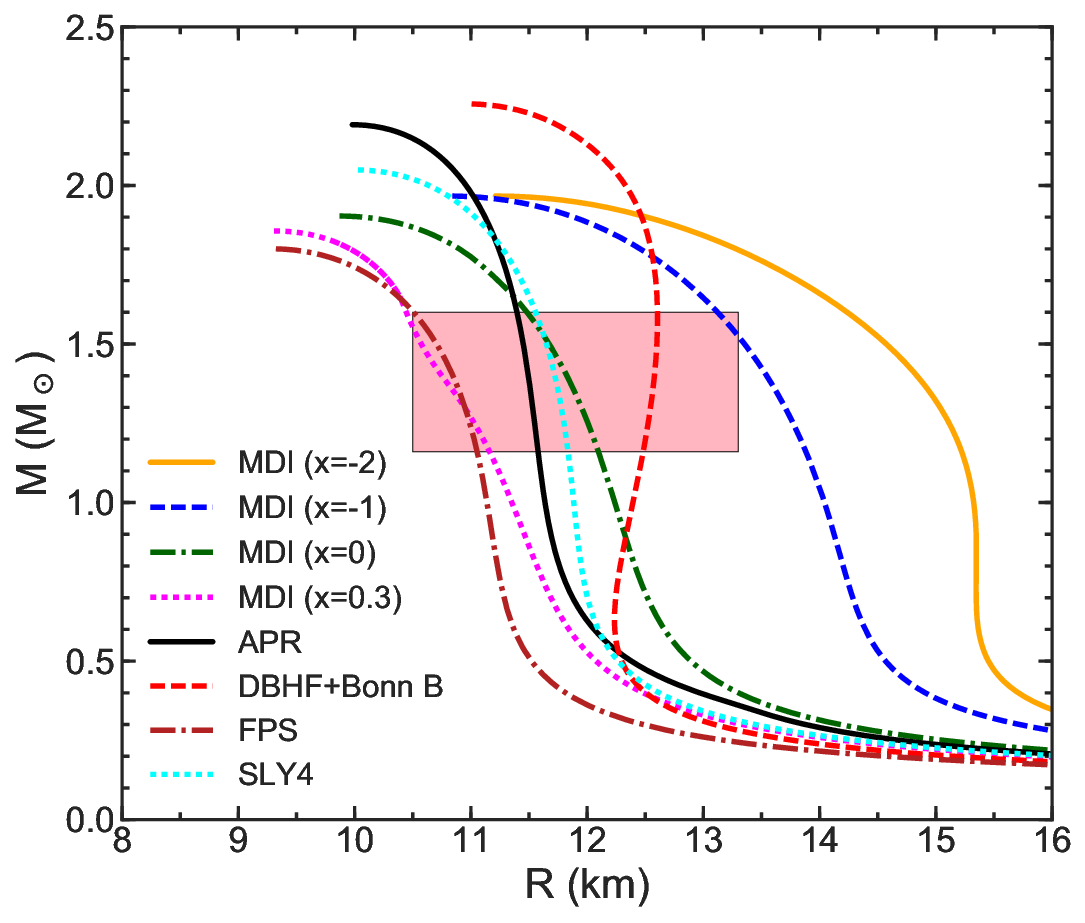}
\caption{(Color online) Neutron star mass-radius relation for the EOSs applied in this study. The shaded region enclosed
between 1.16 and 1.6 $M_{\odot}$ and 10.5 and 13.3 km corresponds to the constraints on the neutron-star radius from
the refined analysis of the GW170817 observation \cite{Abbott:2018exr}.} \label{fig4}
\end{figure}

The details of the EOS can affect significantly global properties of neutron stars and tidal interactions in
inspiraling binary systems. Fig.~\ref{fig4} shows the neutron star mass-radius relation for the EOSs considered
in this study.

\begin{figure}[t!]
\centering
\includegraphics[scale=0.5]{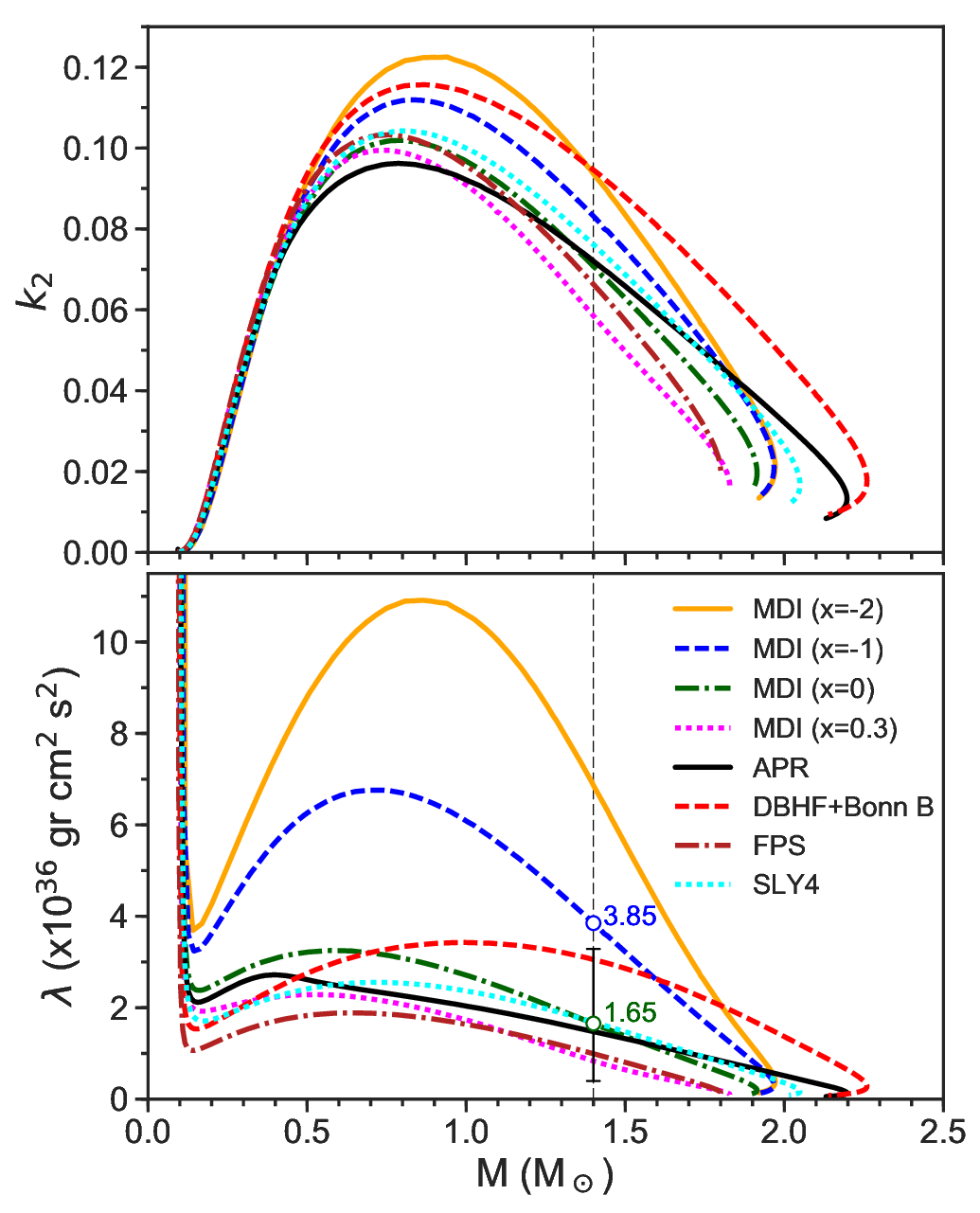}
\caption{(Color online) Love number (upper panel) and tidal deformability (lower panel) as a function
of neutron star mass. The error bar at 1.4$M_{\odot}$ corresponds to the limits on $\lambda_{1.4}$ determined
from the refined analysis of GW170817 \cite{Abbott:2018exr}. The numerical values next to the "o" characters
denote $\lambda_{1.4}$, in units of ($10^{36}$ gr cm$^2$s$^2$), obtained with the MDI EOS with $x=0$ and $x=-1$.} \label{fig5}
\end{figure}

The generic behavior of the Love number $k_2$ is shown in the upper panel of Fig.~\ref{fig5} as a function of stellar mass.
The value of $k_2$ increases rapidly with the neutron star mass approaching a maximum for models with
masses near $1M_{\odot}$, and then gradually decreases. We observe that there is a moderate variation in $k_2$ between
results with different EOSs for fixed neutron star mass. This behavior can be understood in terms of the
physical significance of the Love number $k_2$ -- it measures how easily the bulk of the matter in a neutron star is
deformed~\cite{Hinderer:2009ca}. This implies that more centrally condensed stellar models have smaller Love numbers,
and smaller tidal deformation. On the other hand, for smaller compactness, the softer crust becomes a greater fraction
of the star, so the star is more centrally condensed and $k_2$ becomes smaller.

Next, we turn our attention to the behavior of the tidal deformability $\lambda$. This parameter is proportional to the
quantity that is directly measurable by gravitational-wave observations of inspiraling neutron-star binaries~\cite{Hinderer:2009ca},
and as such has a direct astrophysical significance. We show the tidal deformability in the lower panel of Fig.~\ref{fig5}
as a function of the neutron star mass. Except for very low values of the neutron star mass, for each EOS $\lambda$
follows a trend very similar to that of $k_2$. However, because in addition to the Love number, $\lambda$ is also proportional
to $R^5$ and the applied EOSs result in neutron star configurations with a wide range of radii (see Fig.~\ref{fig4}),
it experiences much greater variations compared to $k_2$. The tidal deformability becomes large for neutron star models with
mass near $0.1M_{\odot}$ because they have large radii. We also observe that $\lambda$ becomes larger for stellar configurations
with masses around 0.6--1.0$M_{\odot}$, where it shows greater variations for EOSs that produce models with large radii. Here we
recall that the neutron star radius is strongly correlated with the density dependence of nuclear symmetry energy~\cite{LP01}.
As a quantitative example, it was demonstrated in Ref.~\cite{Li:2005sr} that changing the parameter $x$ from 0 to -2 while keeping the
incompressibility fixed at $K_0=211$ MeV leads to a change in radius from about 12 km to 15 km for a canonical neutron star without
affecting the maximum neutron star mass ($1.9M_{\odot}$) that can be supported by the MDI EOS. Moreover, the $E_{sym}(\rho)$
of the MDI EOS  with $x=0$ is about the same as that of the APR EOS up to approximately $4\rho_0$. However, the APR EOS has a stiffer
incompressibility of $K_0=269$ MeV. By examining the predictions for the mass-radius relation, it was found in Ref.~\cite{Li:2005sr},
and also shown in Fig. \ref{fig4} in this work, that the APR EOS leads to about a 16\% higher maximum mass ($2.2M_{\odot}$) but only
a 5\% decrease in radius (from 12.0 km to 11.5 km) for a canonical neutron star as compared to the MDI results with $x = 0$.
More qualitatively, an EOS with stiffer $E_{sym}(\rho)$, such as the MDI EOS with $x=-1$ and $x =-2$, results in less centrally
condensed stellar models, and in turn greater radii. Specifically, the MDI ($x=-2$) EOS produces neutron star configurations with larger
radii than those of models from the rest of the EOSs applied here~\cite{KLW2}. Because $\lambda$ quantifies the neutron star
deformation in response to an external tidal field, the results in Fig.~\ref{fig5} suggest that less compact stars are more
easily deformed, and more centrally condensed models are more ``resistant" to deformation. This is consistent with previous
studies which found out that more compact neutron star models are less altered by various deformation driving mechanisms,
e.g., rapid rotation~\cite{KLW2,FPI:1984}.

\begin{figure}[t!]
\centering
\includegraphics[scale=0.43]{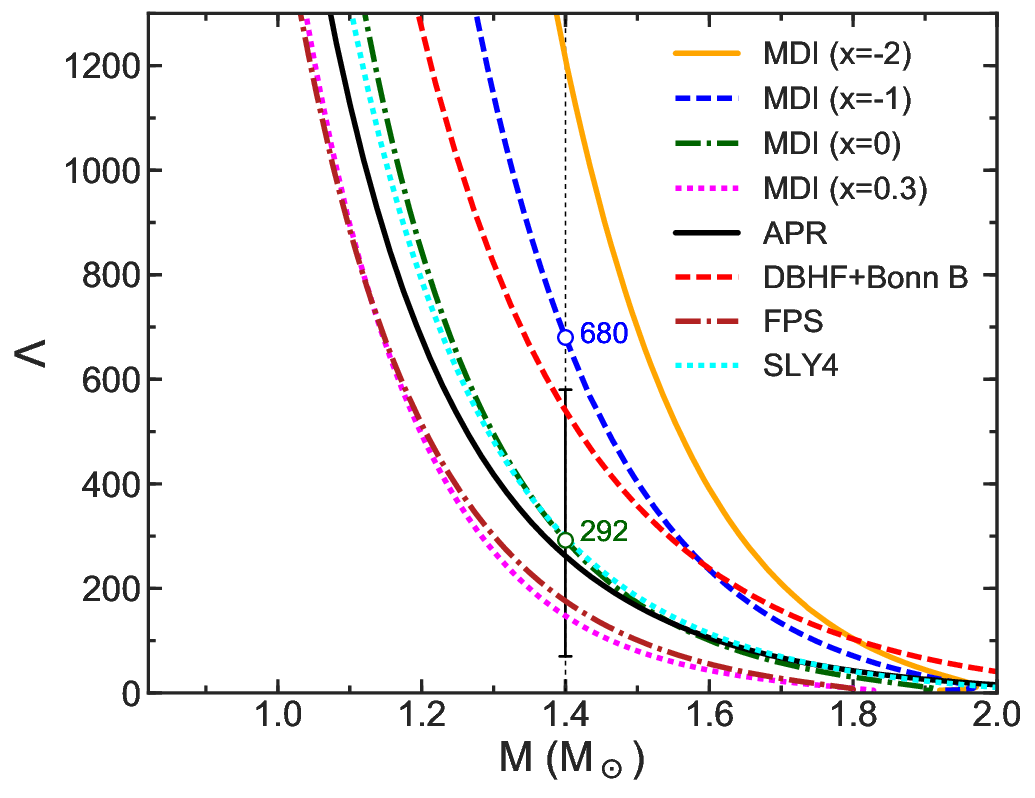}
\includegraphics[scale=0.43]{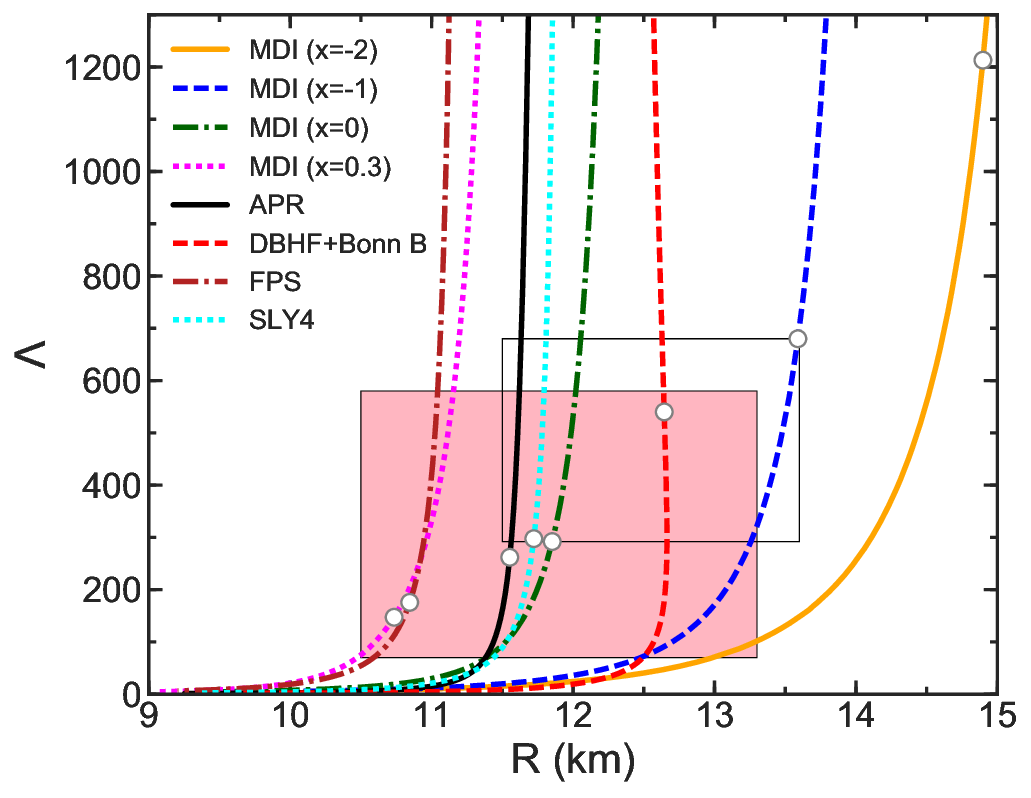}
\vspace{-1mm}
\caption{(Color online) Dimensionless tidal deformability as a function of neutron-star mass (left panel)
and neutron-star radius (right panel). The error bar at $1.4M_{\odot}$ in the left window and the
colored (light-pink) region of $\Lambda=[70 - 580]$ and $R=[10.5 - 13.3]$ km in the right window indicate
the  gravitational wave constraints from GW170817 \cite{Abbott:2018exr}. Similarly, the constraints based
on heavy-ion collisions, obtained with the MDI EOS with $x=0$ and $x=-1$, are denoted by the numerical
values next to the "o" characters in the left panel, and the rectangular region of
$\Lambda=[292 - 680]$ and $R=[11.5 - 13.6]$ km in the right panel. The dot symbols in the right frame
indicate $\Lambda_{1.4}$.} \label{fig6}
\end{figure}

The results in Fig.~\ref{fig5} suggest that $\lambda$ depends on the EOS of stellar matter where the dependence is greater for
stellar models with stiffer symmetry energy, and more generally, stiffer EOSs. As already pointed out in previous
investigations, e.g., Ref. \cite{Fattoyev:2012uu}, for neutron stars in the mass-range of interest, $\lambda$ is mostly affected by
the high-density behavior of $E_{sym}(\rho)$. The symmetry energy has been partially constrained by available terrestrial
nuclear laboratory data, in particular around the saturation density. Although at higher densities $E_{sym}(\rho)$
is presently rather uncertain, these constraints could still provide valuable information about the EOS of dense matter and related
neutron star properties. Since the MDI ($x=0$) and MDI ($x=-1$) EOSs have symmetry energy consistent with nuclear laboratory data
they provide limits on the most probable neutron star configurations, and in turn on the most probable values of $\lambda$,
in accord with the available terrestrial nuclear laboratory constraints. In this respect, the region between the $x=0$ and $x=-1$ curves
in the lower panel of Fig. \ref{fig5} denotes the range of the most probable values of the tidal deformability $\lambda$ in
equal-mass binaries ($\tilde{\lambda}$ reduces to $\lambda$ in the equal-mass case). The specific numeric values shown in the figure
are for neutron star configurations of $1.4M_{\odot}$. Depending on the details of the EOS, $\lambda$ is found to be in the range of
$\sim[1.7 - 3.9]\times 10^{36}$ (gr cm$^2$s$^2$). These estimates, based on nuclear laboratory data, are consistent with the
constraints on the tidal deformability by the LIGO/Virgo observation of the GW170817 event \cite{Abbott:2018exr}.
In the case of a low dimensionless spin ($\chi \leq |0.05|$) of the individual binary components, for $1.4M_{\odot}$ stellar
configurations, Ref.~\cite{Abbott:2018exr} quotes $\Lambda_{1.4}$ in the range of $\sim [70 - 580]$, or
$\lambda \sim [0.4 - 3.3]\times 10^{36}$ (gr cm$^2$s$^2$), where $\lambda=C_{\lambda}\Lambda$ with  $C_{\lambda}=G^4 M^5 / c^{10}$
(in CGS units). These refined constraints from GW170817 are denoted by the error bar at 1.4$M_{\odot}$ in Fig. \ref{fig5}.

\begin{table}[t!]
\caption{Properties of $1.4M_{\odot}$ neutron star models for the EOSs discussed in the text.
The first column identifies the equation of state. The remaining columns exhibit the following
quantities: neutron-star radius $R$ (km); compactness parameter $\beta$; Love number $k_2$;
dimensionless tidal deformability $\Lambda$; slope of nuclear symmetry energy at
saturation density $L$ (MeV).}
\begin{center}
\begin{tabular}{lc@{\hskip 4mm}c@{\hskip 4mm}c@{\hskip 4mm}c@{\hskip 4mm}cc}\hline\hline\label{tab.1}
EOS         &  $R$ & $\beta$ & $k_2$    & $\Lambda$  & $L$\\
\hline
MDI ($x=-2$)      & 14.9     &  0.14  &  0.0936  &  1212   & 153   \\
MDI ($x=-1$)      & 13.6     &  0.15  &  0.0831  &   680   & 107   \\
MDI ($x=0$)       & 11.9     &  0.17  &  0.0707  &   292   &  62   \\
MDI ($x=0.3$)     & 10.7     &  0.19  &  0.0585  &   147   &  48   \\
APR               & 11.5     &  0.18  &  0.0721  &   261   &  62   \\
DBHF+Bonn B       & 12.6     &  0.16  &  0.0946  &   540   &  69   \\
FPS               & 10.8     &  0.19  &  0.0664  &   175   &  35   \\
SLY4              & 11.7     &  0.18  &  0.0762  &   297   &  47   \\
\hline\hline
\end{tabular}
\end{center}
\end{table}

In Fig. \ref{fig6} we show the dimensionless tidal deformability $\Lambda$ as a function of the neutron star mass (left panel)
and stellar radius (right panel). This allows for a direct comparison with the observational constraints from the GW170817
event. Corresponding to the MDI EOS with $x=0$ and $x=-1$, $\Lambda$ is found to be in the range of 292 $\leq\Lambda_{1.4}\leq$ 680.
The error bar at 1.4$M_{\odot}$ in the left window and the colored rectangular region in the right window represent
the tighter constraints from the GW170817 observation \cite{Abbott:2018exr}. The rectangular region of $\Lambda=[292 - 680]$ and
$R=[11.5 - 13.6]$ km represents the constraints on $\Lambda$ and $R$ from heavy ion collision data. As explained in Ref. \cite{Li:2005sr}
the minimum radius is extended to 11.5 km (from 11.9 km as obtained with the MDI ($x=0$) EOS) to account for the remaining
uncertainty in the symmetric part of the EOS. While both regions, based on the GW170817 analysis and heavy-ion
collision data, reasonably overlap, the constraints from nuclear laboratory data appear to be more restrictive. For completeness,
in Table~\ref{tab.1} we list properties of $1.4M_{\odot}$ neutron-star models calculated with all EOSs applied in this work.

\subsection{Effects of varying the density dependence of $E_{sym}$ on the tidal deformability}

To investigate further the effects of $E_{sym}(\rho)$ on $\Lambda$ in the following we study how varying
the density dependence of the symmetry energy affects $\Lambda_{1.4}$. In particular, we examine how
transitions from stiff to soft and soft to stiff $E_{sym}(\rho)$ at saturation density $\rho_0$ impact
the tidal properties of neutron stars. This approach is mainly motivated by the ongoing efforts to pin down
the close but not precisely established relation between $\Lambda_{1.4}$  and $R_{1.4}$, and determine what
aspects of the EOS of dense neutron-rich matter could be exactly  determined by accurate measurements of
the tidal deformability. Although some interesting indications and speculations have been examined in recent
literature, e.g., correlation between $\Lambda_{1.4}$ and the slope of the symmetry energy $L_0$ around $\rho_0$,
sizes of neutron-skins in heavy nuclei, and/or possible phase transitions in dense neutron-rich matter \cite{Zhou:2017pha,Fattoyev:2017jql,Zhang:2018vrx,Zhang:2018vbw,Bhat:2018erd}, much more work is still required to 
determine what aspects of the EOS could be revealed from $\Lambda_{1.4}$. Most notably, the fact that latest 
measurements of  neutron star radii from low-mass X-ray binaries favor softer $E_{sym}(\rho)$ (and EOS)
\cite{Gandolfi:2016,Steiner:2012xt}, while  the measured skin thickness of $^{208}Pb$ by the PREX experiment
\cite{Abrahamyan:2012gp,Horowitz:2012tj} (if confirmed by PREX-II) suggests stiffer $E_{sym}$ at nuclear
densities, led some studies to point out that the evolution from soft to stiff $E_{sym}$ may be indicating
a phase transition in the neutron star interior just above saturation density \cite{Fattoyev:2017jql}.

\begin{figure*}[t!]
\centering
\includegraphics[scale=0.5]{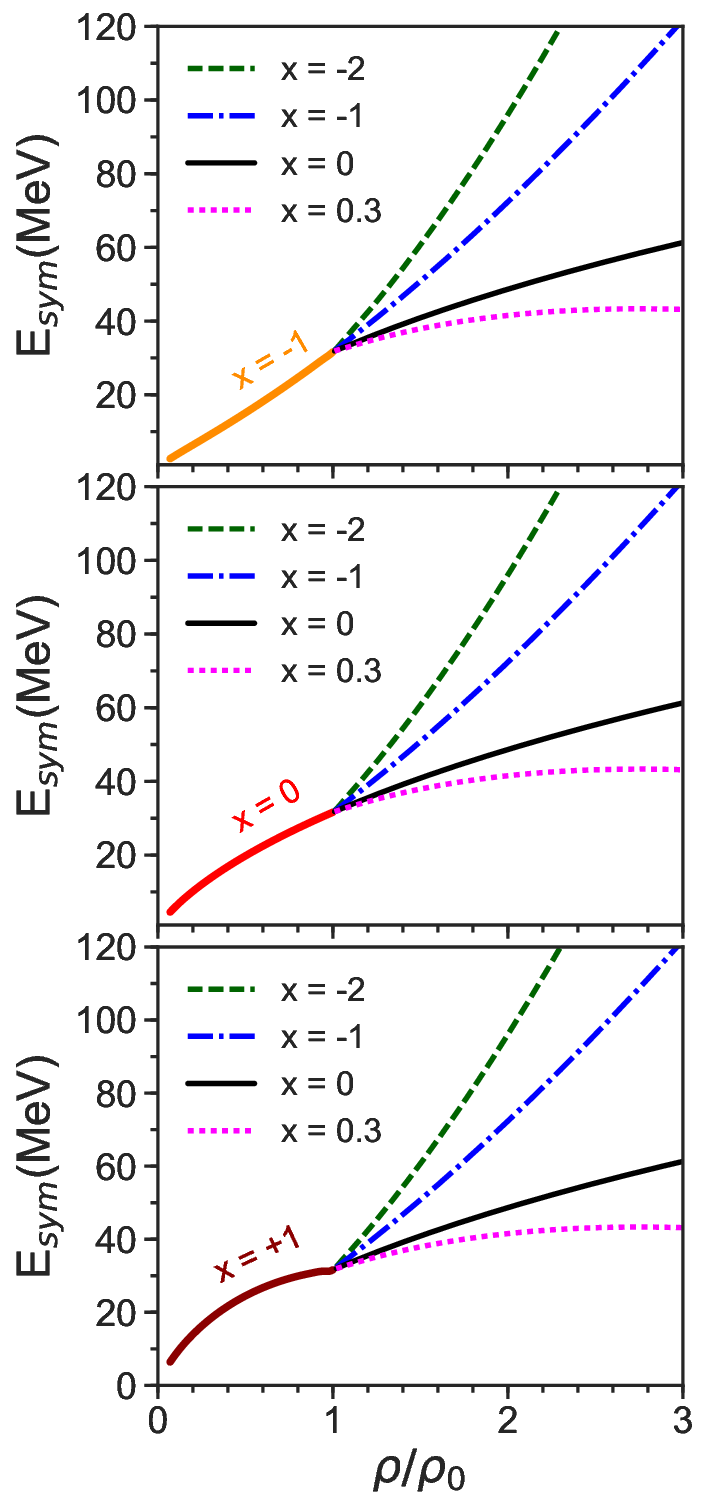}
\includegraphics[scale=0.5]{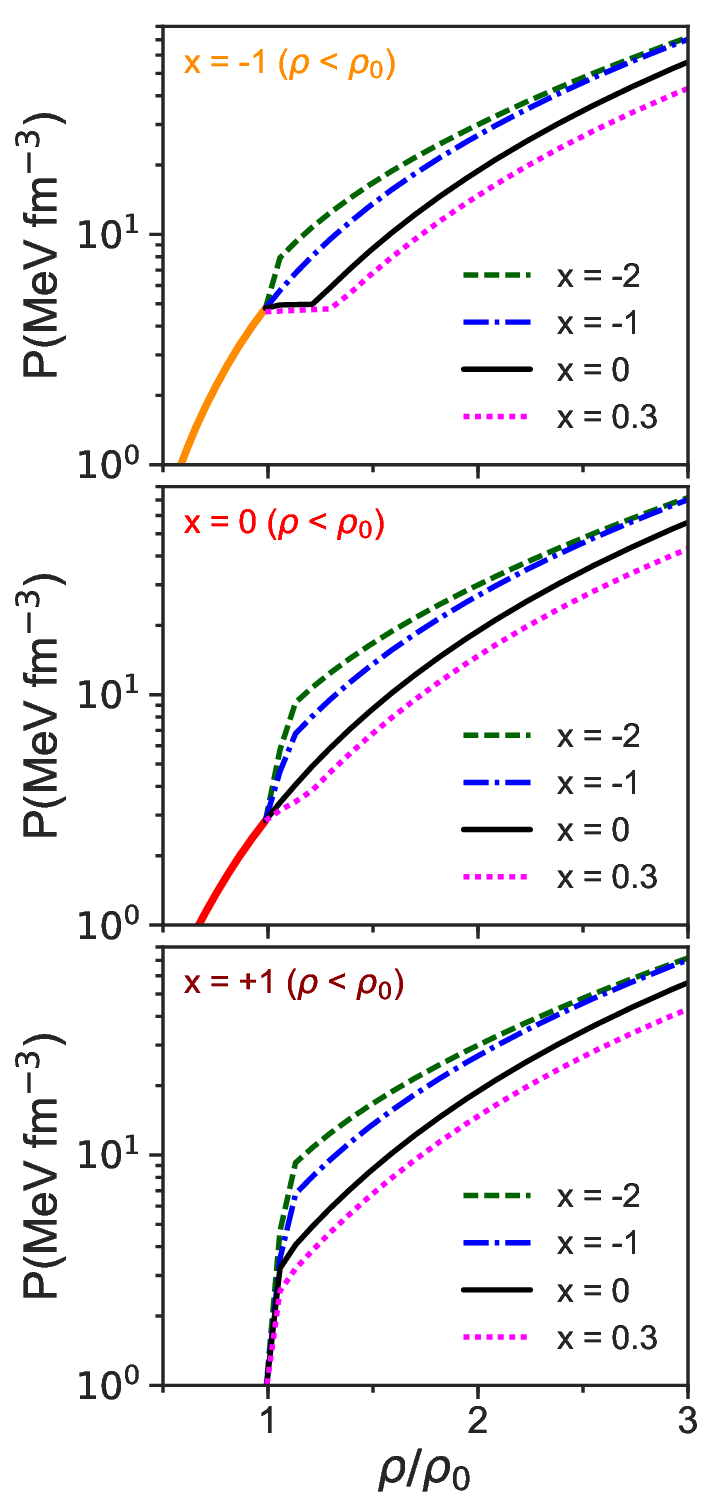}
\caption{(Color online) Varied symmetry energy (left panel) and pressure (right panel). $E_{sym}(\rho)$
and $P$ are shown for three representative values of $x$ below saturation density $\rho_0$: -1 (upper windows),
0 (middle windows), and +1 (lower windows). Above $\rho_0$ $x$ is varied from -2 to 0.3. Specifically, for
$\rho>\rho_0$ we consider $x$ = [-2, -1, 0, 0.3]. See text for details.} \label{fig7}
\end{figure*}

In the MDI model the density dependence of the symmetry energy is controlled by the parameter $x$ and, as illustrated
in Fig. \ref{fig1}, different values of $x$ lead to a wide range of possible trends for $E_{sym}(\rho)$. For the
purpose of this investigation, we model the transition from one trend of the density dependence of $E_{sym}$ to another by
using different values of $x$ below and above $\rho_0$. Namely, for each of several representative values of $x$ below
the saturation density, $x$ = [-1, 0, +1], we vary $x$ above $\rho_0$ from -2 to 0.3 and compute $\Lambda$ with the
resultant EOSs. We refer to $E_{sym}$ with $x$ = -2 and -1 as stiff and $E_{sym}$ with $x$ = [0, 0.3, +1] as soft. Or
more qualitatively, for the purpose of our analysis we define $E_{sym}(\rho)$ as stiff for $K_{sym}>0$ and soft for
$K_{sym}<0$, where as illustrated in Fig. \ref{fig2}, with increasing $x$, $K_{sym}$ changes sign at approximately
$x=-0.5$. Both branches of $E_{sym}$, below and above saturation density, are joined as smoothly as possible at $\rho_0$.
The modified $E_{sym}(\rho)$ and the resultant total pressure $P$ for these cases are shown in Fig. \ref{fig7}.
For instance, in this respect changing $x$ from -1 for $\rho<\rho_0$ to 0 or 0.3 for $\rho>\rho_0$ represents
a transition from stiff to soft symmetry energy (upper panels of Fig. \ref{fig7}). Conversely, changing $x$ from +1
for $\rho<\rho_0$ to -1, or -2, for $\rho>\rho_0$ represents a transition from soft to stiff symmetry energy (lower
panels of Fig. \ref{fig7}). Here we recall that both $L$ and $K_{sym}$ change simultaneously with $x$
(see Fig. \ref{fig2}). Keeping $x$ fixed below $\rho_0$ and varying it above the $\rho_0$ is effectively
equivalent to keeping either $L$ or $K_{sym}$ fixed and varying the other. A systematic study of the effects
of varying the symmetry energy parameters $L$, $K_{sym}$, and the skewness $J_{sym}$ has been performed recently
in Ref. \cite{Zhang:2018vbw}. Instead, in this work we investigate specifically the impact on $\Lambda$ of a
transition from one density dependence of $E_{sym}$ to another (e.g., stiff-to-soft or soft-to-stiff), and assess
whether precise measurements  of the tidal deformability could distinguish such a transition.

\begin{figure*}[t!]
\centering
\includegraphics[scale=0.4]{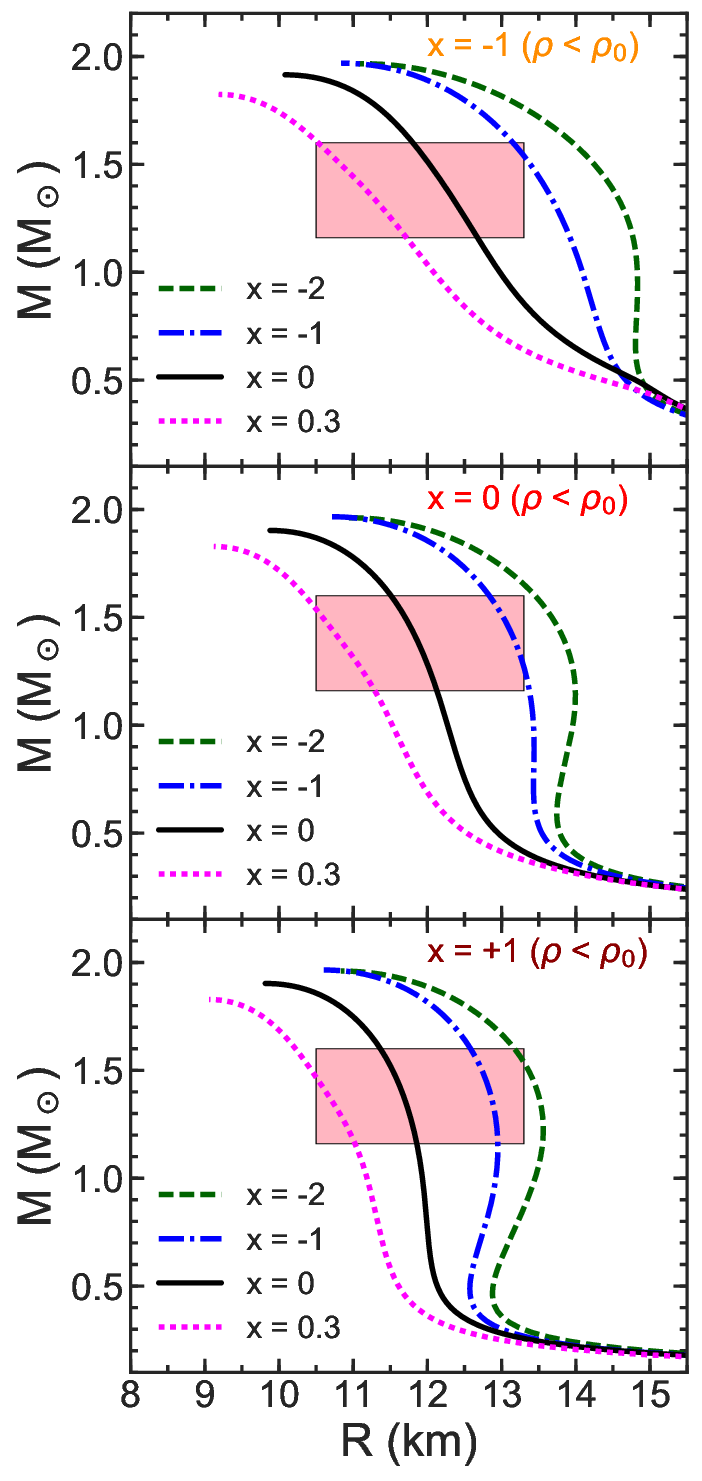}
\includegraphics[scale=0.4]{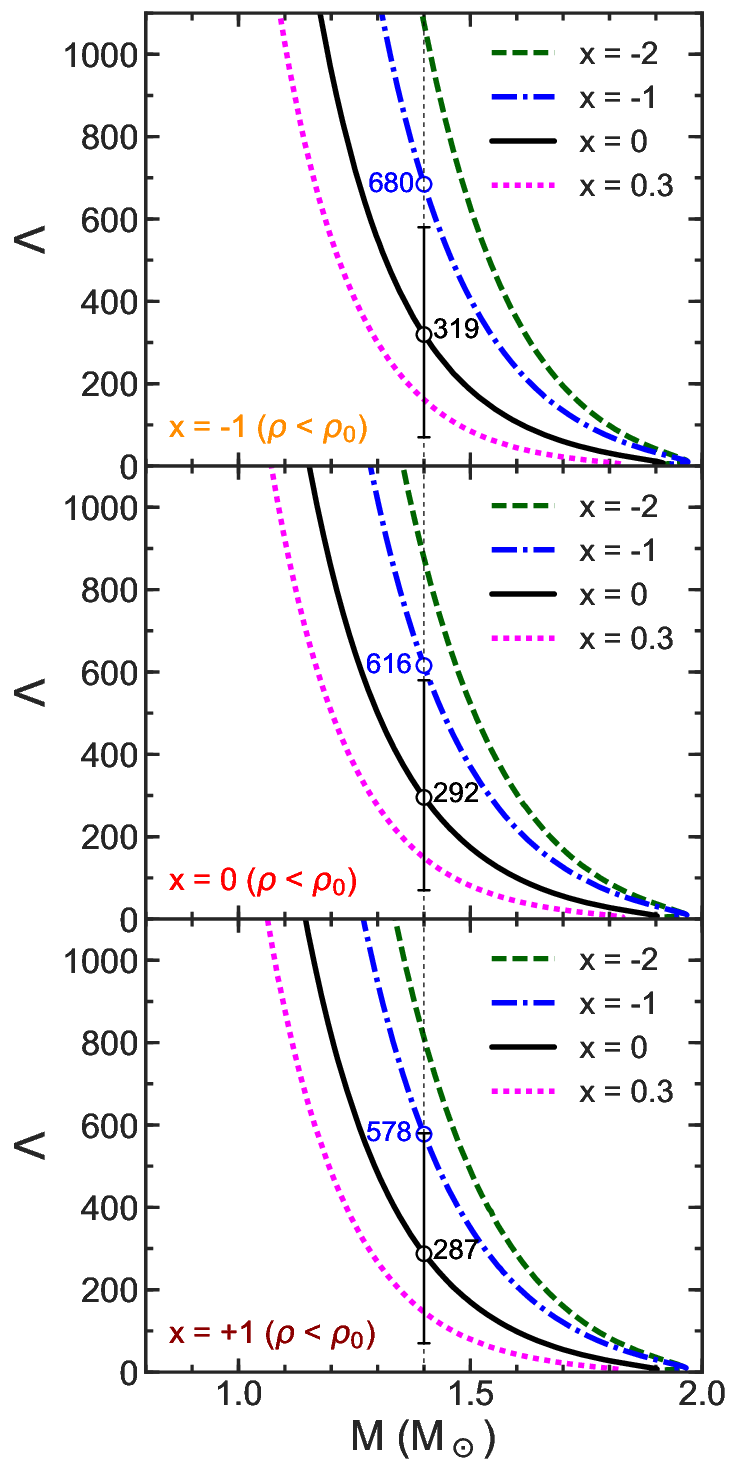}
\includegraphics[scale=0.4]{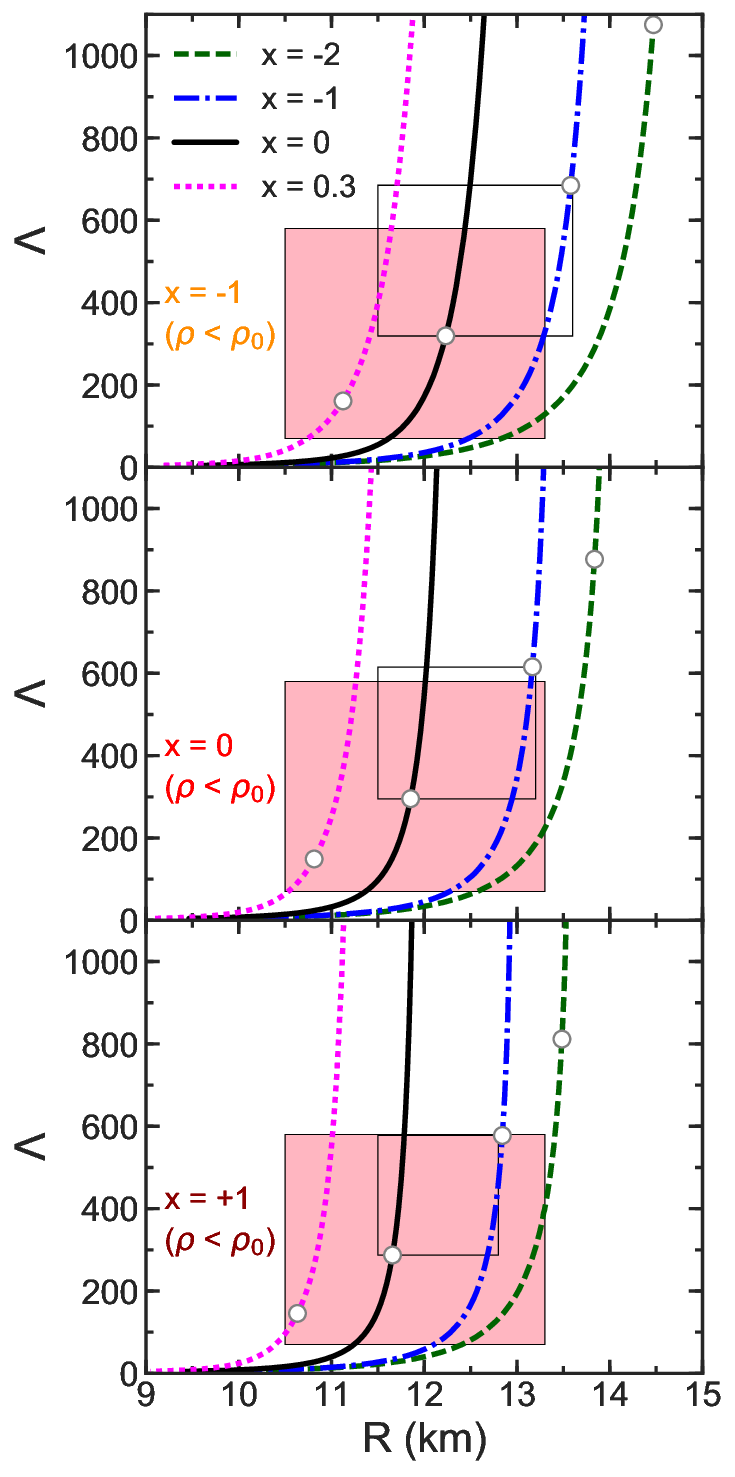}
\caption{(Color online) Mass-radius relation (left panel) and dimensionless tidal deformability
$\Lambda$ as a function of neutron-star mass $M$ (middle panel) and radius $R$ (right panel).
In all frames, the curve labels indicate $x$ values above saturation density ($x_{(\rho>\rho_0)}$).
Colored regions (left and right windows) and the error bar at 1.4$M_{\odot}$ (middle window)
represent the gravitational-weave constraints from GW170817 \cite{Abbott:2018exr}. Uncolored rectangular
regions in the right panel indicate $R_{1.4}$ and $\Lambda_{1.4}$ computed with the MDI EOS with
$x_{(\rho>\rho_0)}=0$ and $x_{(\rho>\rho_0)}=-1$. As explained in the text and Ref. \cite{Li:2005sr}
the minimum radius is extended to, and kept at, 11.5 km. The "o" characters (middle and right
windows) denote values of $\Lambda_{1.4}$. See text for details.} \label{fig8}
\end{figure*}

The results of this analysis are displayed in Fig. \ref{fig8}, where we show the mass-radius relation (left panel) and
$\Lambda$ as a function of the neutron-star mass $M$ (middle panel) and radius $R$ (left panel), and Table \ref{tab.2}
which summarizes properties of $1.4M_{\odot}$ neutron star models. We examine how the transition form stiff to soft
and soft to stiff  symmetry energy affects the tidal deformability. First, we start with the stiff-to-soft
$E_{sym}$ transition. Specifically, we consider the case where for $x=-1$ ($L=107$, $K_{sym}=+95$) for
$\rho<\rho_0$ and $x=0$ ($L=62$, $K_{sym}=-87$) for $\rho>\rho_0$. We see that $R_{1.4}$ decreases from 13.6
km to 12.2 km, $k_{2,1.4}$ from 0.0831 to 0.0661 and $\Lambda_{1.4}$ from 680 to 319, while compactness $\beta$
increases from 0.15 to 0.17. Similarly, for the case of $x_{(\rho<\rho_0)}=-1$ and $x_{(\rho>\rho_0)}=0.3$
$R_{1.4}$ decreases to 11.6 km, $k_{2,1.4}$ to 0.0537 and $\Lambda_{1.4}$ to 161, while $\beta$ increases to
0.19. More generally, such qualitative behavior is observed in all instances where $x_{(\rho<\rho_0)}<x_{(\rho>\rho_0)}$.
Next, we look at the soft-to-stiff $E_{sym}(\rho)$ transition, where as a representative example we consider the
case of $x_{(\rho<\rho_0)}=0$ and $x_{(\rho>\rho_0)}=-1$. We observe that $R_{1.4}$ increases from 11.9 km to
13.2 km, $k_{2,1.4}$ from 0.0707 to 0.0882, and $\Lambda_{1.4}$ from 292 to 615, while $\beta$ decreases from 0.17
to 0.16. This trend is seen in all cases where $x_{(\rho<\rho_0)}>x_{(\rho>\rho_0)}$. For the stiff-to-soft
$E_{sym}$ transition we also notice that although $R$, $k_2$ and $\Lambda$ shift toward their canonical values
(as summarized in Table \ref{tab.1}), the actual decreased values are always larger than the canonical ones.
Conversely, for the soft-to-stiff $E_{sym}$ transition, the increased values of $R$, $k_2$, and $\Lambda$ tend
to be smaller than their canonical counterparts. The observed behavior is easily understood, and best interpreted,
in terms of the modified total pressure (left panel of Fig. \ref{fig7}). For instance, in the case of a
stiff-to-soft $E_{sym}$ transition the dips exhibited by the $x=0$ and $x=0.3$ curves (in the upper left
window of Fig. \ref{fig7}) indicate a decrease in $P$, as compared to the $x=-1$ curve for $\rho<\rho_0$,
and result in overall softer EOSs and more centrally condensed neutron-star models with smaller $R$ and
$\Lambda$. By the same token, in the case of a soft-to-stiff $E_{sym}$ transition, the humps exhibited by the
$x=-1$ and $x=-2$ curves (in the middle left window of Fig. \ref{fig7}) represent an increase in $P$, as compared
to the $x=0$ curve for $\rho<\rho_0$, and lead to less compact NS configurations with larger $R$ and $\Lambda$.
In addition, the corresponding decreased (increased) values of $R$ and $\Lambda$ are larger (smaller) than their
canonical counterparts because the resultant modified EOSs used to compute the NS properties are overall
stiffer (softer) than the unaltered EOSs (computed with the same $x$ for the whole density range).

\begin{table}[t!]
\caption{Properties of $1.4M_{\odot}$ neutron star (NS) models with the MDI EOS with varied $E_{sym}(\rho)$
as discussed in the text. NS properties are given for three representative values of $x$ for $\rho<\rho_0$:
-1, 0, and +1. For each $x$ below $\rho_0$, the parameters $L$ and $K_{sym}$ are also included. The
first column indicates the value of the parameter $x$ below saturation density. The
remaining columns exhibit the following quantities: $x$ for $\rho > \rho_0$; compactness parameter $\beta$;
NS radius $R$ (km); Love number $k_2$; dimensionless tidal deformability $\Lambda$.}
\begin{center}
\begin{tabular}{c|c@{\hskip 1mm}c@{\hskip 4mm}c@{\hskip 4mm}c@{\hskip 3mm}cc}\hline\hline \label{tab.2}
$x_{(\rho<\rho_0)}$   &  $x_{(\rho>\rho_0)}$   &  $\beta$ & $R$ & $k_{2}$   & $\Lambda$\\
\hline
-1               &  -2  &   0.14 &  14.5 & 0.0960 &  1074 \\
$L=107$          &  -1  &   0.15 &  13.6 & 0.0831 &   680 \\
$K_{sym}=95$     &   0  &   0.17 &  12.2 & 0.0661 &   319 \\
                 & 0.3  &   0.19 &  11.1 & 0.0537 &   161 \\
\hline
0                &  -2  &   0.15 &  13.8 & 0.0980 &   876 \\
$L=62$           &  -1  &   0.16 &  13.2 & 0.0882 &   615 \\
$K_{sym}=-87$    &   0  &   0.17 &  11.9 & 0.0707 &   292 \\
                 & 0.3  &   0.19 &  10.8 & 0.0571 &   149 \\
\hline
+1               &  -2  &   0.15 &  13.5 & 0.1032 &   811 \\
$L=48$           &  -1  &   0.16 &  12.8 & 0.0937 &   577 \\
$K_{sym}=-142$   &   0  &   0.18 &  11.7 & 0.0756 &   287 \\
                 & 0.3  &   0.19 &  10.6 & 0.0607 &   145 \\
\hline\hline
\end{tabular}
\end{center}
\end{table}

It is also seen in Fig. \ref{fig8} that as $x_{(\rho<\rho_0)}$ increases from -1 (upper frames) to 0 (middle frames),
to +1 (lower frames), $R$ and $\Lambda$ decrease and this effect is best observed in the left panel. It is signified
by the diminishing area of the uncolored rectangular region defined by $R_{1.4}$ and $\Lambda_{1.4}$ computed with
the MDI EOS with $x=-1$ and $x=0$ above $\rho_0$. This behavior is attributed to the overall softening of the EOS
due to the increase of $x$ below saturation density. Most importantly, it is also observed that the region
representing the MDI EOS results (uncolored rectangle) overlaps significantly with the colored rectangular region
denoting the gravitational wave constraints, with the overlap being complete in the lower left frame of Fig. \ref{fig8}.
These findings suggests that even if such a low-density phase transition exists, it would not be visible in the current
GW170817 data. They also illustrate that a variety of density behaviors of the symmetry energy, as parameterized here
by $x_{(\rho<\rho_0)}$ and $x_{(\rho>\rho_0)}$, could yield the same $\Lambda$. More qualitatively, since a single
$x$ value corresponds to specific values of $L$ and $K_{sym}$ (Fig. \ref{fig2}), varying $x$ below and above
saturation density is equivalent to varying the slope and curvature of $E_{sym}$, and many different combinations of
large (small) slope and small/negative (large/positive) curvature lead to the same $\Lambda_{1.4}$. In this respect,
a precise measurement of $\Lambda_{1.4}$ restricts the configuration space of possible $L$ and $K_{sym}$,
but not the individual parameters of $E_{sym}$ themselves. These conclusions are consistent with recent
investigations \cite{Zhang:2018vbw} where the effects on $\Lambda_{1.4}$ of varying $L$, $K_{sym}$ and
$J_{sym}$ are studied systematically with a parameterized EOS. There does not appear to exist a simple
relationship between the tidal deformability and the slope of the symmetry energy (hence the neutron-star radius
$R$) alone, and the reported relations are rather model dependent \cite{Zhu:2018ona}. As already pointed out
in the literature, translating $\Lambda$ measurements directly into $R$ constraints has to be taken with
some caution \cite{Zhu:2018ona}. Therefore, both $R$ and $\Lambda$ have to be measured independently to pin down
the details of $E_{sym}(\rho)$.

It is expected, in the near future, that many more observations of GWs from merging neutron star binaries with the
advanced ground based and scheduled spacebourne gravitational-wave detectors will become available, and allow for
precise $\Lambda$ measurements. Such tighter constraints on the tidal deformability together with precise
measurements of the neutron-star radius from X-ray observations will determine the exact correlation between $R$ and
$\Lambda$ and help further understand the high-density $E_{sym}$ and EOS.

\section{Summary and outlook}

In summary, using the MDI EOS which has its SNM part and the low-density symmetry energy $E_{sym}(\rho)$ constrained
by earlier data from heavy-ion reactions, we have investigated the effect of the symmetry energy $E_{sym}(\rho)$ on
the tidal properties of coalescing binary neutron stars. For $1.4M_{\odot}$ neutron-star models, the dimensionless
tidal deformability $\Lambda_{1.4}$ is found to be in the range of $\sim[292 - 680]$. These estimates based on nuclear
laboratory data are in excellent agreement with the constraints on $\Lambda_{1.4}$ from the GW170817 event
\cite{Abbott:2018exr}. The GW170817 event provides a limit for the radius of canonical neutron stars. It is consistent
with but less restrictive than the earlier prediction based on the EOS partially constrained by the experimental data
of heavy-ion reactions. We also investigated the effect of varying $E_{sym}(\rho)$, below and above $\rho_0$, on
tidal properties of neutron stars and found that while $\Lambda_{1.4}$ is rather sensitive to this variation,
many different combinations of large (small) $E_{sym}$ slope and small/negative (large/positive) $E_{sym}$ curvature
yield the same $\Lambda_{1.4}$. Specifically, we studied the transition from stiff to soft $E_{sym}$ at $\rho_0$ and
our findings suggest that even if such a low-density transition exists it would not be visible in the current GW170817
data. Therefore, additional probes including ones from terrestrial nuclear laboratories are required resolve
this degeneracy and determine the precise density behavior of $E_{sym}$. In this respect, the results in this
work signify the importance of coherent analyses of dense neutron-rich nuclear matter EOS underlying both
nuclear laboratory experiments and astrophysical observations.

The gravitational wave astronomy is at its very beginning but it has already started providing important insights about
the nature of compact stars and the EOS of dense neutron-rich matter. In the near future, besides the ongoing and planned
X-ray observations of neutron stars, more neutron star merger events, together with a new  generation of gravitational-wave
detectors and refined waveform models, will certainly help us better understand the structure of compact stars and the nature
of dense neutron-rich matter. Similarly, new experiments in terrestrial nuclear laboratories, especially at advanced rare
isotope beam facilities, will also provide greater details about the EOS of neutron-rich matter, in particular about the
high-density behavior of the nuclear symmetry energy. A truly multi-messenger approach combining results from both terrestrial
laboratories and astrophysical observations will finally allow us to pin down the EOS of neutron-rich matter in a broad
density range.

\ack{We would like to thank F. J. Fattoyev, N.B. Zhang and J. Xu for helpful communications and
discussions. This work is supported in part by the U.S. Department of Energy, Office of Science,
under Award Number DE-SC0013702, the CUSTIPEN (China-U.S. Theory Institute for Physics with Exotic
Nuclei) under the US Department of Energy Grant No. DE-SC0009971, the National Natural Science
Foundation of China under Grant No. 11320101004 and the Texas Advanced Computing Center.}

\end{document}